\documentclass[
    ,final            
  ]
  {aipproc}
\layoutstyle{6x9}
\usepackage{amsmath,amssymb}
\usepackage{latexsym}

\begin{document}
\newcommand{\Tr}{\mathrm{Tr}}
\newcommand{\tr}{\mathrm{tr}}
\newcommand{\Det}{\mathrm{Det}\;}
\newcommand{\Pexp}{\mathrm{Pexp \,}}
\newcommand{\point}{\quad .}
\newcommand{\comma}{\quad ,}
\newcommand{\D}{\mathcal{D}}
\newcommand{\Dslash}{\ensuremath \raisebox{0.025cm}{\slash}\hspace{-0.32cm} D}
\newcommand{\Aslash}{\ensuremath \raisebox{0.025cm}{\big\slash}\hspace{-0.25cm} A}
\newcommand{\dslash}{\not{\hbox{\kern-2pt $\partial$}}}
\newcommand{\be}{\begin{equation}}
\newcommand{\ee}{\end{equation}}
\newcommand{\beq}{\begin{eqnarray}}
\newcommand{\eeq}{\end{eqnarray}}
\title{Spontaneous breaking of discrete symmetries in QCD on a small volume}


\author{B. Lucini\footnote{Presenter at the conference.}~}{
  address={Department of Physics, Swansea University, Singleton Park, Swansea SA2 8PP, UK}
}

\author{A. Patella}{
  address={Scuola Normale Superiore, Piazza dei Cavalieri 27, 56126 Pisa, Italy
\\and INFN Pisa, Largo B. Pontecorvo 3 Ed.~C, 56127 Pisa, Italy}
}

\author{C. Pica}{
  address={Physics Department, Brookhaven National Laboratory, Upton, NY 11973, USA}
}

\begin{abstract}
In a compact space with non-trivial cycles, for sufficiently small values of
the compact dimensions, charge conjugation ($\mathcal{C}$), spatial reflection
($\mathcal{P}$) and time reversal ($\mathcal{T}$) are spontaneously broken in
QCD. The order parameter for the
symmetry breaking is the trace of the Wilson line wrapping around the compact
dimension, which acquires an imaginary part in the broken phase. We show
that a physical signature for the symmetry breaking is a persistent baryonic
current wrapping in the compact directions. The existence of such a current
is derived analytically at first order in perturbation theory and confirmed
in the non-perturbative regime by lattice simulations.
\end{abstract}
\maketitle
\section{Breaking of discrete symmetries}
We consider a non-Abelian gauge theory with gauge group SU($N$) and
$N_f$ families of fundamental Dirac fermions. This theory is described
by the Lagrangian
\beq 
\label{l0}
\nonumber
{\cal L} =  - \frac{1}{2 g^2} \mbox{Tr}\left( G_{\mu \nu} (x) G^{\mu \nu}(x) \right) + \sum_{n=1}^{N_f} \bar{\psi}_n (x)\left(i \dslash - \Aslash(x) - m \right) \psi_n(x) \ , 
\eeq
where $G_{\mu \nu} = \partial_{\mu} A_{\nu} - \partial_{\nu} A_{\mu} + [A_{\mu},A_{\nu}]$.\\
The manifold on which the theory is defined is $T \times L^3$, where $T$ corresponds to time and the compact spatial directions $L$ are equal. Spatial directions are closed with periodic boundary conditions (PBC), while the temporal direction is closed with antiperiodic boundary conditions for fermions and PBC for bosons.\\
We compute the vacuum expectation value ({\em vev}) of the Wilson line in a compact direction~\cite{Unsal}
\begin{equation}
W^{(A)}_{\alpha} (x, z^{(\alpha)}) = \mathrm{Pexp} \; \left( i \int_0^{L} A_\alpha (x,z^{(\alpha)}) \, d z_\alpha \right) \ .
\end{equation}
The gauge can be fixed in such a way that $W$ is diagonal 
\beq
W^{(A)}_\alpha (x, z^{(\alpha)}) = \left( e^{iv^{(A)}_{1 \alpha}(x, z^{(\alpha)})}, \dots , e^{iv^{(A)}_{N \alpha}(x, z^{(\alpha)})} \right) \ .
\eeq
We focus on the effective potential for	those eigenvalues\footnote{ We drop the subscript $\alpha$ where this does not lead to ambiguities.}:
\begin{eqnarray}
e^{i T V(v_1, \dots , v_N)} &=&
\int \, \exp \left\{ i \int {\cal L} d^4 x \right\} \times \\
\nonumber
&& \times \prod_k \delta \left( v_k - \frac{1}{T L^2} \int v^{(A)}_k(\mathbf{x}) \, dt dx_{\perp} \right)
\, \D A \D \bar{\psi} \D \psi \ .
\end{eqnarray}
The absolute minima of $V$ $(v_1, \dots , v_N)$ give the expectation values for the set of the eigenvalues. If $\mathcal{C}$-symmetry is broken, the set of eigenvalues is not invariant under the substitution $v \rightarrow -v$. The effective potential for the Wilson line is given by~\cite{Hoyos}
\begin{eqnarray}
\label{ppot}
&& V(\vec{v}_1, \dots, \vec{v}_N) = \left[ \sum_{i,j = 1}^{N} f(0, \vec{v}_i-\vec{v}_j) - 2 N_f \sum_{i=1}^{N} f(m, \vec{v}_i) \right] \ ,\\
&& \mbox{with} \ f(m,\vec{v}) =  \frac{1}{L} \left( \frac{m L}{\pi}\right)^2 \sum_{\vec{k} \neq 0} \frac{K_2( m L k)}{k^2} \sin^2 \left( \frac{1}{2} \vec{k}\cdot \vec{v} \right)
 \ ,
\end{eqnarray}
where $K_2$ is the modified Bessel function of the second kind of order 2 and the sum runs over three-index integer vectors $\vec{k}$. The first term of Eq.~(\ref{ppot}) gives rise to an attraction between the eigenvalues. The second term produces an unconstrained absolute minimum at $v_{i} = \pi$. When the $SU(N)$ constraint is taken into account, the minima are:
\begin{equation}
v_{1}^* = v_{2}^* = \dots = v_{N}^* = \left\{
\begin{array}{l}
\pm \frac {N-1}{N} \pi \qquad \mbox{for $N$ odd}\\
\pi \qquad \qquad \ \ \mbox{for $N$ even}\\
\end{array}
\right. \ .
\end{equation}
The spontaneous symmetry breaking shows up as a non-zero imaginary part of the Wilson line. Hence, if $N$ is even, there is no symmetry breaking. If $N$ is odd, then $\mathcal{P}$, $\mathcal{C}$, $\mathcal{T}$, $\mathcal{CPT}$ are broken. There are $8$ vacua. The same calculation can be reproposed for a number $n$ of compact dimensions. We arrive at the same conclusions, with the number of vacua given by $2^n$. This result is due to the spacial PBC for the fermions. The validity of the calculation in the non-perturbative regime has been checked on the lattice in~\cite{DeGrand}.
\section{Symmetry breaking and physical observables}
The Wilson line is an order parameter for the symmetry breaking.
However, if the theory is not invariant under ${\cal P}$,
${\cal C}$ and ${\cal T}$ we expect this to be reflected by some
observable. This observable must be odd under the broken
symmetries, but even under the combined action of two of them.
The spatial components of the baryon current
$j_\alpha = \langle \sum_{i=1}^{N_f} \bar{\psi}_i \gamma_\alpha
\psi_i \rangle$ satisfy this property.
Why should we expect a non-zero baryon current? A non trivial {\em vev} of
the Wilson line means a non-zero value of the gauge field in that direction.
Since the system is translationally invariant along the compact direction,
the value of the gauge field must be constant (note that in the presence of
toroidal topology a constant field cannot be gauged away). The background
gauge field acts as a non-trivial source for the baryon current. Hence,
we expect this current to be different from zero.\\
The expectation value of the baryonic current is generated by the partition function of the system in presence of a generalised "chemical potential", which acts as a source for the current~\cite{Lucini}. Using the one-loop value for the effective potential we obtain
\begin{eqnarray}
\langle \vec{j} \rangle = - \frac{N_f N}{L^3} 
\left( \frac{m L}{\pi}\right)^2 \sum_{\vec{k} \neq 0} \frac{K_2( m L k)}{k^2} \sin \left(\vec{k}\cdot \vec{v}^* \right) \vec{k} \ .
\end{eqnarray}
$\langle j_{\alpha} \rangle$ is zero when $v_{\alpha}^* = 0$ or  $v_{\alpha}^* = \pi$ (i.e. when the symmetry breaking does not occur in direction $\alpha$), is odd under $v_{\alpha}^* \to -v_{\alpha}^*$, goes to zero when $m \to \infty$.\\
\section{The lattice calculation}
\begin{figure}
       \includegraphics[scale=0.4]{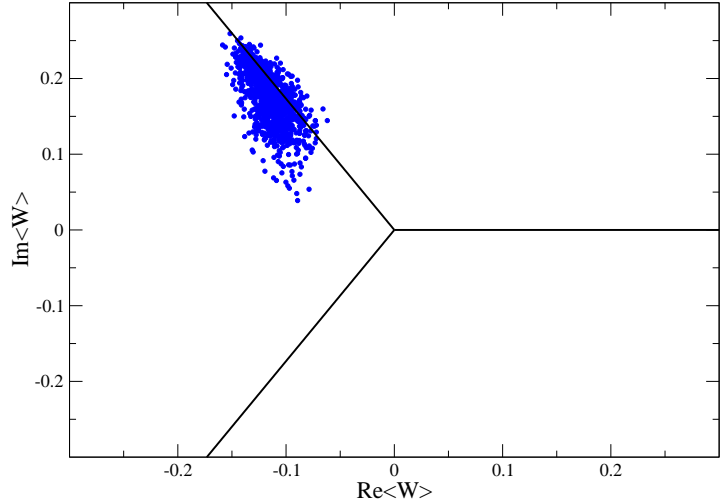}
       \label{wloopplot}
       \caption{Plot of the Wilson loop. Clustering of the phase around $2 \pi/3$ means spontaneous breaking of discrete symmetries.}
\end{figure}

\begin{figure}
  \label{figcurrent}
  \includegraphics[scale=0.4]{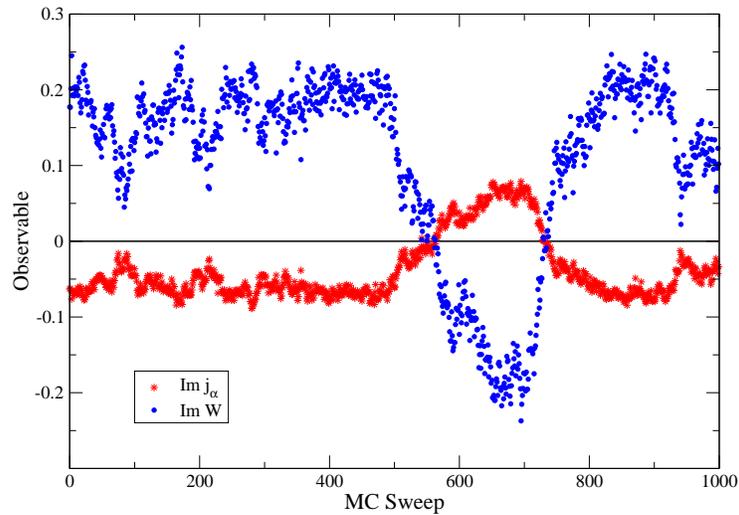}
  \caption{Monte Carlo history of the imaginary part of the Wilson line and the current in a compact direction in the broken symmetry phase.}
\end{figure}

Details of the lattice action used in our simulation are provided in~\cite{Lucini}. A $24 \times 4^3$ lattice at $\beta = 5.5$ was used. This fixes the lattice spacing (determined measuring the Sommer parameter $r_0$~\cite{Sommer}) to 0.125 fm. This means that $L_t  = a N_t = 3$ fm and $L_s = a N_s = 0.5$ fm,
in agreement with the assumptions of sufficiently large $L_t$ and $L_s$ below
the fermi scale. The breaking of discrete symmetries can be checked by looking at the Wilson line wrapping around a spatial direction. A typical behaviour is
plotted in Fig.~\ref{wloopplot}, which shows that the Wilson loop magnetises along $e^{i \frac{2}{3} \pi}$, i.e. $\langle W \rangle$ acquires an imaginary part, as required by the symmetry breaking scenario.\\
For the Euclidean rotated theory in the broken phase, we expect a non-zero value of the imaginary part of the current. At the given lattice parameters we find
$|\mbox{Im}\langle j_{\alpha} \rangle| = 0.060 \pm 0.002 $, which should be compared with the perturbative prediction $\langle j_{\alpha} \rangle \simeq 0.037473(4)$: it is remarkable that for compact dimensions of the order of $1/\Lambda_{QCD}$ the perturbative prediction still gives the correct order of magnitude.\\The correlation of the current with the phenomenon of the spontaneous symmetry breaking can be understood by looking at the correlations between Im$j_{\alpha}$ and Im$W$. An example is given in Fig.~\ref{figcurrent}.
This plot shows that the two observables are closely related. In particular,
in the event of a tunneling between the broken vacua (possible in our case because the spatial volume is finite), the current changes sign.
\section{Conclusions and outlook}
In the phase in which discrete symmetries are spontaneously broken there is a persistent baryonic current wrapping around the topologically non-trivial compact directions. Unlike persistent currents in superconductors, this current is conserved. The existence of this current is a clear physical signature of the symmetry breaking and could be used to determine the order of the symmetry restoring phase transition that happens at a critical radius of the compact direction, which string-inspired calculations predict to be second order~\cite{Armoni:2007jt}. Another open problem is the interplay between the aforementioned phase transition and the chiral restoring phase transition at finite temperature.

\end{document}